# QUANTUM SUPERPOSITION PRINCIPLE REFORMULATION AND COLLAPSE OF WAVE FUNCTION EXPLANATION

I. G. KOPRINKOV

Department of Applied Physics, Technical University of Sofia,
8 Kl. Ochridski Blvd, Sofia 1000, Bulgaria
*E-mail*: igk@tu-sofia.bg

*Abstract*. The quantum superposition principle is reexamined and reformulated based on the adiabatic theorem of quantum mechanics, nonadiabatic dressed states and experimental evidences. The collapse of the wave function and the quantum measurement problem are explained within the reformulated quantum superposition principle.

*Key words*: quantum superposition, adiabatic theorem, dressed states, wave function, collapse.

## 1. INTRODUCTION

The quantum superposition principle is one of the fundamental principles of quantum mechanics. It states that if two or more quantum states $|\psi(\vec{r},t)\rangle_i$ are states of a given quantum system, *i.e.*, they obey the respective Schrödinger, any linear combination of these states is also a state of same quantum system:

$$|\psi(\vec{r},t)\rangle = \sum_i c_i(t) |\psi(\vec{r},t)\rangle_i. \qquad (1)$$

The quantum superposition principle, Eq. (1), is correct from a formal mathematical point of view due to the linearity of the Schrödinger equation. The following understandings are assigned to the conventional quantum superposition, Eq. (1): the superposition of quantum states is *simultaneous*, *coherent* and may involve *any kind of states* of the quantum system. The simultaneity, however, has never been proved experimentally. The typical characteristic time of electron motion in atoms falls in the attosecond time range, 1 as = $10^{-18}$ s. The current attophysics and technology may reach this time range but the exact timing of the superimposed quantum states is not yet tested experimentally. The coherence is closely revealed to the ability of the quantum system to interfere. The quantum states can interfere if the quantum system is simultaneously in these states. Thus, the simultaneity has a crucial importance for the quantum superposition. Finally, the superimpisition may involve any kind of states, most frequently – the stationary eigenstates of the quantum systems. In addition, the physical mechanism of the superposition is not revealed assuming that it is a natural consequence from the physical nature of the quantum phenomena.

In this work we show that the eigenstates of stationary or adiabatic Hamiltonians cannot be superimposed simultaneously and, in this way, they cannot interfere. This invokes a reexamination of the conventional quantum superposition principle, Eq. 1. It is done based on the adiabatic theorem of quantum mechanics [1], experimental evidences [2, 3] and nonadiabatic dressed states [4]. A new formulation of the quantum superposition principle is proposed. Physical states that satisfy the specified requirements for quantum superposition are found. Explanation of some basic quantum mechanical problems as the collapse of wave function and the quantum measurement problem become a natural consequence from the reformulated quantum superposition principle.

## 2. PHYSICAL ARGUMENTS IN THE REFORMULATION OF THE QUANTUM SUPERPOSITION PRINCIPLE

Our approach to the quantum superposition principle is based on:
1. General theoretical argument – adiabatic theorem of quantum mechanics [1].
2. Special theoretical argument – nonadiabatic dressed states [4].
3. Experimental evidences [2, 3].

### 2.1. ADIABATIC THEOREM OF QUANTUM MECHANICS

The adiabatic theorem of quantum mechanics, first formulated in [1], states that a quantum system remains in an instantaneous eigenstate $\psi(t)$ of its Hamiltonian $\hat{H}(t) = \hat{H}_0 + \hat{H}'(t)$, i.e., $\hat{H}(t)\psi(t) \equiv (\hat{H}_0 + \hat{H}'(t))\psi(t) = E(t)\psi(t)$, if it changes slow enough, i.e., adiabatically, due to a given perturbation $\hat{H}'(t)$, and if there is a gap [5] between the energy of this state and the rest part of the Hamiltonian spectrum. When the perturbation terminates, the quantum system will be in the same quantum state $\psi_0$, from which the adiabatic evolution begins, and no transition to other state will occur. Consequently, according to the adiabatic theorem, *a quantum system cannot be simultaneously in more than one eigenstate of adiabatically changing Hamiltonian*.

The adiabatic theorem can also be applied to the stationary eigenstates, bare states (BS), $\psi_0$ of nonperturbed Hamiltonian $\hat{H}_0$ of a completely isolated (closed) quantum system, $\hat{H}_0 \psi_0 = E_0 \psi_0$. The BS are quantum states at perfect adiabatic conditions as, at lack of perturbation $\hat{H}'(t) = 0$, the total Hamiltonian is not simply adiabatic but it is even stationary, $\hat{H}(t) \equiv \hat{H}_0 = $ const. Adiabatic theorem in this perfect adiabatic case states that if a closed quantum system is in given BS of its Hamiltonian $\hat{H}_0$, it will remain in that BS and no transition to other BS will occur.

Hence, *a closed quantum system cannot be simultaneously in more than one stationary eigenstate, BS, of its Hamiltonian $\hat{H}_0$*.

In conclusion from the adiabatic theorem: (*i*) the eigenstates of adiabatic or stationary Hamiltonians cannot be simultaneously superimposed, as it is supposed in Eq. (1); (*ii*) the physical reasons for transitions between quantum states are nonadiabatic factors acting on the quantum system.

### 2.2. NONADIABATIC DRESSED STATES

Three generations of quantum states can be distinguished: bare states, adiabatic (dressed) states (ADS) [2, 3] and nonadiabatic dressed states (NADS) [4]. The BS are states of a closed quantum system. The ADS are states of quantum system in presence of classical adiabatic electromagnetic field. The NADS are states of quantum system in presence of nonadiabatic electromagnetic field and environment. The NADS represent a generalization of ADS and BS. The following notations for ground and excited BS, ADS, NADS will be used: $|g\rangle$ and $|e\rangle$, $|G\rangle$ and $|E\rangle$, $|\tilde{G}\rangle$ and $|\tilde{E}\rangle$, respectively, which stand for all characteristics of the states (quantum numbers, symmetries, etc.).

Each NADS consists of real and virtual components/states [4]:

$$|\tilde{G}\rangle = \cos(\theta/2)|\tilde{G}_r\rangle + \sin(\theta/2)|\tilde{G}_v\rangle \tag{2a}$$

$$|\tilde{E}\rangle = \cos(\theta/2)|\tilde{E}_r\rangle - \sin(\theta/2)|\tilde{E}_v\rangle. \tag{2b}$$

The real (index *r*) and the virtual (index *v*) components of NADS (at, *e.g.*, ground state initial conditions) are:

$$|\tilde{G}_r\rangle = |g\rangle \exp\left[-i\int_0^t \tilde{\omega}_G dt'\right] \tag{3a}$$

$$|\tilde{G}_v\rangle = |e\rangle \exp\left[-i\int_0^t (\tilde{\omega}_G + \omega)dt' - i\varphi(t)\right] \tag{3b}$$

$$|\tilde{E}_r\rangle = |e\rangle \exp\left[-i\int_0^t \tilde{\omega}_E dt' - i\varphi(t)\right] \tag{3c}$$

$$|\tilde{E}_v\rangle = |g\rangle \exp\left[-i\int_0^t (\tilde{\omega}_E - \omega)dt'\right] \tag{3d}$$

The real components of NADS, $|\tilde{G}_r\rangle$ and $|\tilde{E}_r\rangle$, of energies (Bohr frequencies) $\tilde{\omega}_G$ and $\tilde{\omega}_E$ (Eqs. (3a) and (3c)) are (quasi)stationary states in which the quantum system exists within the lifetime of the state and originate, after continuous evolution, from the respective BS, $|g\rangle$ and $|e\rangle$, of energies $\omega_g$ and $\omega_e$, respectively (Fig. 1). The virtual components of NADS $|\tilde{G}_v\rangle$ and $|\tilde{E}_v\rangle$ of energies $\tilde{\omega}_G + \omega$ and $\tilde{\omega}_E - \omega$ (Eqs. (3b) and (3d)), are *new states* that originate from the respective real components of NADS, $|\tilde{G}_r\rangle$ and $|\tilde{E}_r\rangle$, respectively, after absorption/emission of one photon from/to the field of energy (frequency) $\omega$ (Fig. 1). The virtual components can be created everywhere on the energy scale depending on photon energy and exist only during the action of the field. The field "lifts up"/"pulls down" the population by one photon energy (for single-photon processes) from the real component of given NADS to the virtual component of same NADS. *The virtual components of NADS are real physical states*, but not simply a mathematical construct [4]. The quantities $\cos(\theta/2)$ and $\sin(\theta/2)$ are probability amplitudes for the real and virtual components, respectively.

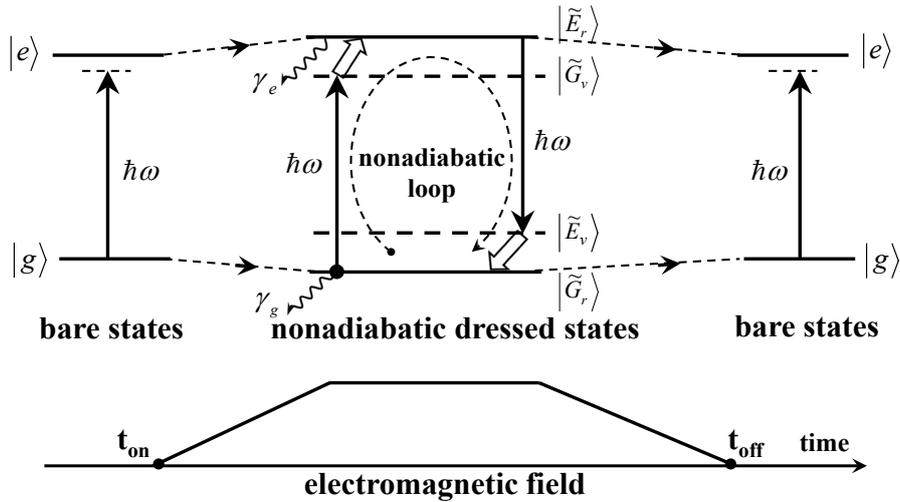

Fig. 1 – Evolution of BS toward NADS and back to BS with switching on/off the electromagnetic field and damping. The real and the virtual components of the NADS are shown with full and broken lines, respectively. The radiative, nonadiabatic and damping processes are presented by solid, hollow and wavy arrows, respectively.

The range of validity of BS, ADS and NADS along with the accompanied physical processes are shown in Fig. 2. The NADS are only the states that can operate directly in the area of action of nonadiabatic factors, *i.e.*, where, in agreement with the adiabatic theorem, the quantum transitions take place (Fig. 2).

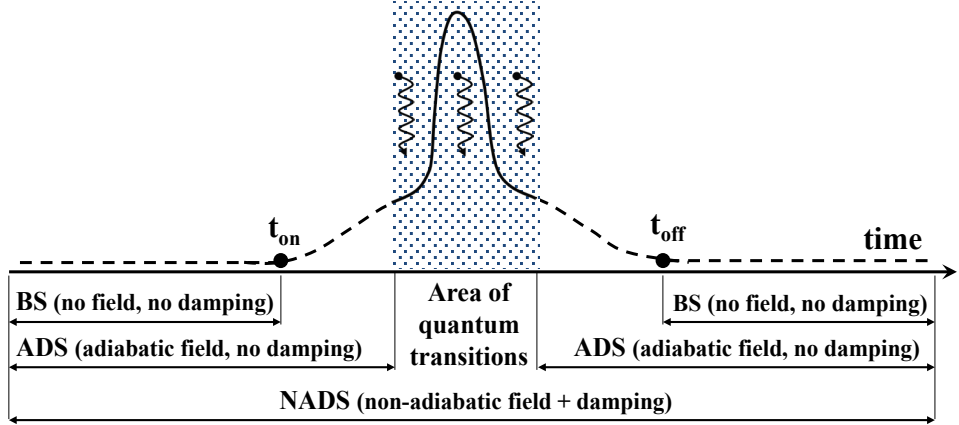

Fig. 2 – The range of validity of BS, ADS and NADS. The broken line shows zero or adiabatic field, the full line shows nonadiabatic field, the wavy lines show damping (nonadiabatic) processes. The shaded area is the area where quantum transitions take place.

### 2.3. EXPERIMENTAL EVIDENCES

Some remarkable experiments [2, 3], while having only nanosecond time resolution, allow to distinguish the population dynamics of real and virtual states using a spectrally resolved pump-probe approach. To get a closer relation to these experiments, the terms and notations of the NADS are applied to the energy levels involved (Fig. 3). Atoms are excited adiabatically by a pump laser pulse at frequency $\omega$ from the real ground NADS $|\tilde{G}_r\rangle$ to the virtual ground NADS $|\tilde{G}_v\rangle$. Two other laser fields at frequency $\omega_{p1}$ and $\omega_{p2}$ in resonance with a high-lying state $|n\rangle$ probe the population on the virtual $|\tilde{G}_v\rangle$ and the real $|\tilde{E}_r\rangle$ state. While the pump pulse is adiabatic, population of the real state $|\tilde{E}_r\rangle$ is observed due to nonadiabatic factors from the environment (collisions with other atoms, mainly).

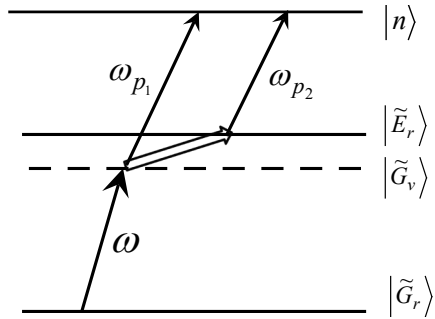

Fig. 3 – Energy levels and quantum transitions diagram to test the real and virtual states [2,3].

The main results from these experiments can be summarized as: (*i*) real population on the virtual state $|\tilde{G}_v\rangle$ is detected and it may exceed by an order of magnitude the population on the real state $|\tilde{E}_r\rangle$; (*ii*) the population on the virtual state $|\tilde{G}_v\rangle$ is proportional to the intensity of the field; (*iii*) the population on the real state $|\tilde{E}_r\rangle$ is proportional to the time integral of the intensity of the field.

The following conclusions about the population dynamics can be done from these experiments taking also into account the nature of the involved physical processes and the features of the real and virtual components of the NADS [4]:

(*i*) the quantum system physically resides on the virtual state $|\tilde{G}_v\rangle$ because a real population is detected on this state [2, 3]. This is the main argument that the virtual states are real physical states but not simply a mathematical construct [4].

(*ii*) the virtual state $|\tilde{G}_v\rangle$ is populated first due to the following arguments: the virtual state $|\tilde{G}_v\rangle$ is created by an ultrafast electron process and its probability amplitude, $\sin(\theta/2)$, becomes nonzero immediately with switching the field on [4]; the real state $|\tilde{E}_r\rangle$ is far off resonance (in this case) and becomes adiabatically decoupled from the virtual state $|\tilde{G}_v\rangle$ for direct radiative population; the population on the virtual state $|\tilde{G}_v\rangle$ strongly exceeds this on the real state $|\tilde{E}_r\rangle$; part of the population on the virtual state $|\tilde{G}_v\rangle$ (high population) is transferred, at some later time, to the real state $|\tilde{E}_r\rangle$ (low population) due to the action of nonadiabatic factors – much slower, in this case, atomic collisions.

(*iii*) the virtual state $|\tilde{G}_v\rangle$ exists only during the action of the electromagnetic field that creates this state and its amplitude, $\sin(\theta/2)$, and thus – the population, follows the intensity of the field. The real state $|\tilde{E}_r\rangle$ exists independently on the field and the population is trapped and accumulated in the real state during the lifetime of the state, which explains the integral time dependence of the population.

## 3. REFORMULATION OF THE QUANTUM SUPERPOSITION PRINCIPLE

As has been shown above, the eigenstates of stationary or adiabatic Hamiltonians cannot be superimposed simultaneously and thus – they cannot interfere.

Then, the following questions naturally arise: (*i*) "Is there any kind of quantum states, which can be superimposed simultaneously?"; (*ii*) "What is the physical mechanism of such a superposition?". The above problems have never been considered, to the best of our knowledge, in relation to the conventional quantum superposition and its reconsideration and reformulation is required. This will be done based the adiabatic theorem of quantum mechanics [1], the features of the NADS [4] and the relevant experimental observations [2, 3].

The two-level analog of the conventional quantum superposition, Eq. (1), is

$$|\psi(\vec{r},t)\rangle = c_g(t)|g\rangle + c_e(t)|e\rangle, \qquad (4)$$

where $|g\rangle$ and $|e\rangle$ are real states, most frequently, the BS of the stationary Hamiltonian $\hat{H}_0$, which, according to the adiabatic theorem, cannot be superimposed simultaneously.

We will analyze the NADS from a point of view of the quantum superposition. The two-level NADS also consist of $|g\rangle$ and $|e\rangle$ type of states but only one of these states is real, while the other state is virtual. In particular, $|\tilde{G}_r\rangle$ is a real $|g\rangle$-type state, while $|\tilde{G}_v\rangle$ is a virtual $|e\rangle$-type state; $|\tilde{E}_r\rangle$ is a real $|e\rangle$-type state, while $|\tilde{E}_v\rangle$ is a virtual $|g\rangle$-type state (Eqs. (2), (3)). The virtual component of a given NADS has "opposite" characteristics to these of the real component of the same NADS due to single-photon electric dipole field-matter interaction, considered here. Consequently, *each NADS, ground $|\tilde{G}\rangle$ and excited $|\tilde{E}\rangle$, is a linear superposition of a real and a virtual state* (Eqs. (2), (3)), where the probability amplitudes, $\cos(\theta/2)$ and $\sin(\theta/2)$, stand for superposition coefficients. At zero field, the amplitude of the virtual component is zero, $\sin(\theta/2)$, and the NADS consists of a real component, only, the initial BS from which it originates. Increasing the field amplitude, the amplitude of the virtual component increases while this of the real component decreases and at extremely strong fields they become equal [4]. The virtual component appears at nonzero field irrespectively if the field is adiabatic or nonadiabatic. The virtual component appears simultaneously from the respective real component of given NADS with switching the field on and disappears with switching the field off. The virtual components, $|\tilde{G}_v\rangle$ and $|\tilde{E}_v\rangle$, cannot exist independently on the respective real components, $|\tilde{G}_r\rangle$ and $|\tilde{E}_r\rangle$, from which they originate, and the forcing electromagnetic field. Therefore, we consider (within the concept of quantum state) that the real and virtual component of a given NADS represents a *simultaneous superposition* within the overlap time of the real component lifetime (if not ground state) and the time of action of the field that creates the virtual component.

The formation of simultaneous superposition between real and virtual states within the ground and excited NADS from the ground and excited BS with switching the field on is given by the following expressions:

$$|g\rangle \xrightarrow{E_0 \neq 0} |\tilde{G}\rangle = \cos(\theta/2)|\tilde{G}_r\rangle + \sin(\theta/2)|\tilde{G}_v\rangle \tag{5a}$$

$$|e\rangle \xrightarrow{E_0 \neq 0} |\tilde{E}\rangle = \cos(\theta/2)|\tilde{E}_r\rangle - \sin(\theta/2)|\tilde{E}_v\rangle \tag{5b}$$

Equations 5(a,b) can be considered as *a two-level representation of the reformulated quantum superposition principle* [6]. It can also be extended toward multilevel quantum systems [7]. *The physical mechanism of creation of the NADS is in fact the physical mechanism of creation of a simultaneous superposition of quantum states*. Thus, the superposition of quantum states is a causal physical process having a well-defined physical mechanism but not an accidental linear combination of states. *The above understandings could be extended to other kinds of states created simultaneously under a coherent forcing physical mechanism.*

The reformulated quantum superposition principle is in mutual agreement with the adiabatic theorem of quantum mechanics [1], the properties of NADS [4] and experimental evidences [2, 3]. It can be easily understood for spectrally resolved quantum states but looks not so transparent if a number of closely spaced stationary states fall within spectral bandwidth of the exciting field, *e.g.*, Rydberg states in atoms [8] or vibrational states in molecules [9], thus creating transient quantum states – wave-packets. Although any state $|\psi(\vec{r},t)\rangle$ can be expanded mathematically, as in Eq. (1), on the stationary eigenstate basis, *i.e.* BS, $|\psi(\vec{r},t)\rangle_i$, the quantum system, physically, is always in a single instantaneous state $|\psi(\vec{r},t)\rangle$, but not in the stationary eigenstates $|\psi(\vec{r},t)\rangle_i$ simultaneously.

## 4. THE COLLAPSE OF WAVE FUNCTION AND QUANTUM MEASUREMENT PROBLEM EXPLAINED WITHIN REFORMRLATED QUANTUM SUPERPOSITION PRINCIPLE

The collapse of the wave function takes place when, in general, the quantum system interacts with the environment or when, in particular, it is a subject to measurement. In this case, the superposition wave function collapses onto one of the eigenstates of the Hamiltonian, the *von Neumann's process 1* [10]:

$$\sum_i c_i(t)|\psi(\vec{r},t)\rangle_i \to |\psi(\vec{r},t)\rangle_k \tag{6}$$

The inability to explain the collapse of the wave function onto only one of the superimposed states, the random outcome of the measurement and the associated physical mechanism are fundamental physical problems, which date since the early days of quantum mechanics. It was expected that the quantum measurement problem could be explained within the theory of decoherence [11–13] but, as it has been realized, decoherence can explain the transition from quantum to classical but cannot explain the collapse of the wave function on single quantum state.

The collapse of the wave function and the quantum measurement problem can be naturally explained within the reformulated quantum superposition principle and the internal dynamics of the quantum system within the NADS. The collapse of simultaneous superposition of quantum states, *e.g.*, the ground and excited NADS, within the reformulated quantum superposition principle (Eqs. (5a, 5b)), can be expressed by Eqs. (7a), (7b), respectively:

$$|\tilde{G}\rangle \equiv \cos(\theta/2)|\tilde{G}_r\rangle + \sin(\theta/2)|\tilde{G}_v\rangle \xrightarrow{E_0 \to 0} |g\rangle \tag{7a}$$

$$|\tilde{E}\rangle \equiv \cos(\theta/2)|\tilde{E}_r\rangle - \sin(\theta/2)|\tilde{E}_v\rangle \xrightarrow{E_0 \to 0} |e\rangle \tag{7b}$$

Hence, the collapse of wave function takes place if the factor(s) creating the quantum superposition, in our case – the electromagnetic field, is removed. At zero field, the virtual component of the NADS disappears and the NADS collapses on the initial BS, from which it originates. Whereas Eqs. (5a) and (5b) represent formation of ground and excited NADS from the respective ground and excited BS at switching the electromagnetic field on, Eqs. (7a) and (7b) represent destruction, *i.e.*, collapse, of the ground and excited NADS on the respective ground and excited BS at switching the electromagnetic field off. Hence, we consider that *the creation of the quantum superposition and the collapse of the wave function are two opposite directions of the same physical process. The collapse of wave function is a real and causal physical process*, which is a direct consequence from the physical mechanism of creation of quantum superposition. It is too fast (based on the characteristic time of electron dynamics) and, thus, too hard, if not impossible for now, to be traced experimentally.

The random outcome of the quantum measurement can be understood from the internal dynamics of the NADS (Fig. 1), whose central part will be called *nonadiabatic loop*. The nonadiabatic loop consists of two arms, ground state arm and excited state arm (Fig. 4). The ground and excited NADS are adiabatically decoupled if the nonadiabatic factors are negligible. Thus, if the quantum system is initially in the ground BS $|g\rangle$, it will evolve toward ground NADS $|\tilde{G}\rangle$ with switching the field on and it will remain in that state with no transition to the excited NADS $|\tilde{E}\rangle$. Due to the action of nonadiabatic factors, quantum transition from $|\tilde{G}\rangle$ to $|\tilde{E}\rangle$, more particularly, from $|\tilde{G}_v\rangle$ to $|\tilde{E}_r\rangle$ (due to the minimal energy gap), will occur.

The quantum system will be trapped in $|\tilde{E}_r\rangle$ and the population will be accumulated there within the lifetime of the state (in agreement with the experimentally observed integral time dependence [2, 3]). Trapping the quantum system in $|\tilde{E}_r\rangle$ terminates its evolution within the ground NADS. If the action of the electromagnetic field continue, it will start forming virtual state $|\tilde{E}_v\rangle$ from $|\tilde{E}_r\rangle$ and the quantum system will evolve within the excited NADS $|\tilde{E}\rangle$ until the next action of nonadiabatic factors transfer it from $|\tilde{E}\rangle$ to $|\tilde{G}\rangle$, more particularly, from $|\tilde{E}_v\rangle$ to $|\tilde{G}_r\rangle$. This terminates the evolution of the quantum system within the excited NADS and closes the nonadiabatic loop. Depending on the duration of measurement and the physical processes involved in the loop, the measurement may last from a part of the loop to a number of loops. The particular outcome of the measurement depends on which state the quantum system is in at the "end of the measurement". As "end of measurement" we understand (depending on the measurement scheme) the instant of time at which: the field is switched off; the quantum system leaves the area of the field; some auxiliary (probe) field brings the quantum system to a state out of the loop, etc. If just before the end of measurement the quantum system evolves within the ground NADS $|\tilde{G}\rangle$, it will be found in ground BS $|g\rangle$ and energy (Bohr frequency) $\omega_g$ will be measured (Fig. 4). If just before the end of the measurement the quantum system evolves within the excited NADS $|\tilde{E}\rangle$, it will be found in excited BS $|e\rangle$ and energy $\omega_e$ will be measured (Fig. 4). As the nonadiabatic factors have, in general, stochastic character, transition from one to other NADS will have the same stochastic character. Consequently, *the nonadiabatic physical processes within the nonadiabatic loop are responsible for the random outcome of the quantum measurement.*

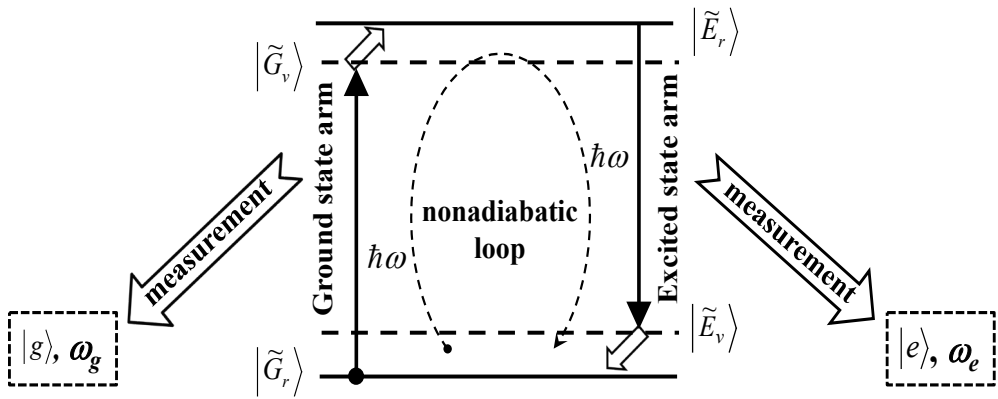

Fig. 4 – Nonadiabatic loop and the random outcome of the quantum measurement.

## 5. CONCLUSIONS

The eigenstates of stationary or adiabatic Hamiltonians cannot be superimposed simultaneously and cannot interfere. The quantum superposition principle is reformulated and the real and virtual components of given nonadiabatic dressed states created by the interaction of the quantum system with a forcing electromagnetic field meet the requirements of quantum superposition. The superposition of quantum states and the collapse of the wave function are found to be two opposite directions of the same physical process. The collapse of the wave function and the quantum measurement problem are explained within the internal dynamics of dressed quantum states.